\documentstyle[twocolumn]{jpsj}

\input psbox.tex

\def\runtitle{Control of Superconducting Correlations in High-Tc Cuprates}
\def\runauthor{Hirokazu {\sc Tsunetsugu} and Masatoshi {\sc Imada}}

\title
{Control of Superconducting Correlations in High-Tc Cuprates
}

\author
{ 
Hirokazu Tsunetsugu and Masatoshi Imada$^{1}$
}

\inst
{
Institute of Materials Physics, University of Tsukuba, 
Tsukuba 305-8573, Japan\\
$^1$Institute for Solid State Physics, University of Tokyo, 
Tokyo 106-8666, Japan
}

\recdate
{
\today
}

\abst
{A strategy to enhance d-wave superconducting correlations is proposed 
based on our numerical study for correlated electron models 
for high-Tc cuprates.  
We observe that the pairing is
enhanced when the single-electron level around 
$(\pi,0)$ is close to the Fermi level $E_F$, while 
the d-wave pairing interaction itself contains
elements to disfavor the pairing due to shift of the
$(\pi,0)$-level.  Angle-resolved photoemission results in
the cuprates are consistently explained in the presence of 
the $d$-wave pairing interaction.  
Our proposal is the tuning of the $(\pi,0)$-level 
under the many-body effects to $E_F$ by optimal design of band structure.}

\kword
{
high-Tc superconductivity, pairing correlations, 
single-electron level, angle-resolved photoemission
}

\begin{document}
\sloppy
\maketitle

We report a new aspect of control of superconducting correlations 
in high-Tc cuprates.  Our new observation 
in the light of numerical studies is about the critical interplay 
between single-electron level at wave vector ($\pi$,0) and (0,$\pi$) 
and electron-pair hopping 
processes, which are usually overlooked.  One potential implication 
is that the superconducting transition temperature $T_c$ 
may be controlled by 
tuning the single-electron level and the pair-hopping coupling constant
in a self-consistent fashion.  
This could provide a guideline for the quest of new materials with 
higher $T_c$.  

Since the first report of high-Tc superconductivity in 1986 \cite{bednorz}, 
various materials with the common CuO$_2$ layer structure have 
been found to show superconductivity with $T_c$ 
up to 133 [K] in mercury compound \cite{schilling} 
and 164 [K] under pressure \cite{gao}.  
One noticeable character common among the high-Tc family is the presence 
of a flat band around the ($\pi$,0) and equivalent (0,$\pi$) points in the 
Brillouin zone \cite{flatband}.  
The anomalous flat dispersion determined by the angular-resolved 
photoemission (ARPES) experiments may be fitted by a $k^4$-type 
form \cite{takahashi}
and numerical calculations on the two-dimensional (2D) Hubbard 
and t-J models 
indicate that it is due to strong electron-electron 
correlations rather than band effects
\cite{dispersion,hubbard,flat1,flat2}.  
ARPES data in fact show the d-wave gap starts growing from the 
flat-dispersion region in underdoped 
samples \cite{underdope1,underdope2,insulator,norman,camp}, 
suggesting a key
role of this region.  However, in the insulating or 
very lightly doped samples, the broad quasiparticle peak
at $(\pi,0)$ appears substantially lower than the level at
$(\pi/2,\pi/2)$ \cite{insulator,calcium1,calcium2,lsco}, 
which implies rapid change in the shape of the Fermi
surface with increasing doping.  Experimentally, the rigid band
picture thus fails and it is yet clear how the electron 
correlation and pairing
interaction effects modify the single-particle dispersion.

In this letter, we first show how the dynamical superconducting 
correlations are enhanced by controlling the single-electron level at 
the wave vector ${\bf k}=(\pi,0)$, 
and we will show later that self-consistent determination of 
the flat-dispersion level around
$(\pi,0)$ under strong correlation effects opens a way to control 
the enhancement.
We employ the 2D t-J model \cite{tjmodel1,tjmodel2,tjmodel3} 
for our numerical 
investigation, and introduce the second and third neighbor 
hoppings in order to control this 
energy level: 
$ H_{tJ}$ = $- \sum_{\sigma}\sum_{i,j} t_{ij} c_{i\sigma}^\dagger 
   c_{j\sigma} $
   $+ J \sum_{\langle i,j \rangle} {\bf S}_{i} \cdot {\bf S}_{i} , 
  \label{tJ}$
where the hopping integrals are $t_{ij}=t$ for the nearest-neighbor pairs, 
$t_1$ and $t_2$ for the second and third neighbor pairs, respectively, 
and otherwise 0.  
The values of $t$ and $J$ are determined from various experiments, 
and a good estimate is $t$=0.3[eV] and $J$=0.12[eV] in 
all high-Tc compounds.  Hereafter, we will use these values 
unless mention explicitly, and measure energy in units of $t$.  
The single-electron kinetic energy is then given 
by $\varepsilon_0 ({\bf k}) $=$-2t ( \cos k_x + \cos k_y ) 
- 4t_1 \cos k_x \cos k_y - 2t_2 (  \cos 2k_x + \cos 2k_y )$, 
and the tight-binding Fermi surface changes its 
shape at half filling from the $\pi/4$-rotated square due to 
the added $t_1$- and $t_2$-terms.  
Fermi surface in various high-Tc compounds has been 
determined by ARPES experiments \cite{flatband,review}, 
and based on comparison with the band calculation: 
for example, for YBa$_2$Cu$_3$O$_7$, see Ref.~\cite{bandcal}.  
A couple of previous studies argue that 
these distant hoppings alone account for the change 
in Fermi surface \cite{distant,calcium2}.  
Another explanation is provided for the deep ($\pi$,0)-level, 
based on the spin-charge separation scenario \cite{spincharge}.  
We take an alternative interpretation on this difference 
to account for superconducting correlations.  

\begin{figure}[tb]
$$ \psboxscaled{300}{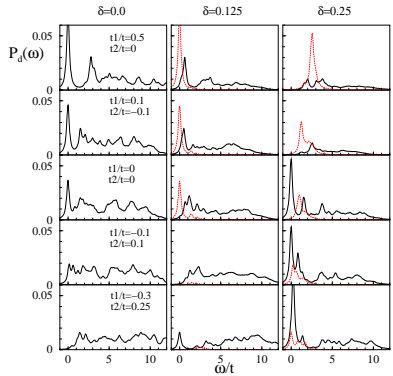} $$ \vspace{-10mm}
\caption{Dynamic correlations of electron-pair annihilation (solid) 
and creation (dotted) with the $d_{x^2-y^2}$ symmetry 
in the 2D t-J model.  $J/t$=0.4.  
}
\label{fig1}
\end{figure}

In order to investigate the enhancement of superconductivity, 
we have calculated its dynamic correlation function, $P_d (\omega)$, 
instead of the usual equal-time correlations.  
It is essential to separate the coherent part to measure 
the enhancement, since only the low-energy dynamics of 
Cooper pairs is relevant to the superconducting 
transition.   
Typical results are shown in Figure~1 for various $t_1$ and $t_2$ 
for creating a spin-singlet hole pair on 
neighboring sites, 
$\Delta \equiv \sum_{j,a} f_d(a) c_{j\uparrow}
c_{j+a \downarrow}$ with 
the $d_{x^2-y^2}$ form factor $f_d(a)=+1$ ($-1$) 
when $a = \pm x$ ($\pm y$).  
This is calculated at zero temperature 
from the ground-state wave function obtained by exact 
diagonalization for the 4$\times$4-site cluster.  
The cluster is chosen such that we can compare the electron 
level between the two characteristic ${\bf k}$-points 
on the Fermi surface at half filling, ($\pi$,0) and ($\pi/2$,$\pi/2$).  
Note that in the noninteracting system, 
the difference in kinetic energy, 
$\varepsilon_0 (\pi,0) - \varepsilon_0 (\pi/2,\pi/2)$, 
is given by $u_0 \equiv 4( t_1 - 2t_2 )$.  

An important result is that upon 
doping a single hole pair into the half-filled insulator 
the coherent peak $P_d (\omega =0) $ 
grows monotonically with increasing $u_0$.  
This is a clear evidence of the evolution 
of coherence of hole-pair motion.  
On the other hand, when $u_0$ is negative 
and large, the coherent peak loses its weight, which means 
the motion of the hole pair is substantially damped.  
We can easily understand this behavior by considering the single-electron 
level.  
For larger $u_0$, $\varepsilon_0 ({\bf k})$ is higher 
around $(\pi,0)$ and the energy cost of 
creating a hole pair is smaller there.  
The $d_{x^2-y^2}$-wave Cooper pairs under consideration consist 
mainly of those holes, and consequently the superconducting 
correlations are enhanced.  
The $u_0$-dependence is no longer monotonic 
when a hole pair is added to the systems already doped, 
as shown in Figure 1 for the doping $\delta$=0.125, 
but a similar interpretation holds.  
For $u_0 >0 $, electrons have been already 
removed from the $(\pi,0)$ and $(0,\pi)$ points at finite dopings, 
and an additional hole pair can 
hardly reside around these ${\bf k}$-points.   
On the other hand, when $u_0 < 0$, electrons are removed 
first from the $(\pm \pi/2, \pm \pi/2)$ and then 
from $(\pi,0)$ and $(0,\pi)$ upon hole doping.  
Therefore, when $u_0$ is negative, the electron occupancy 
at a finite doping may be such that the $(\pi,0)$-level 
is still below and close to the Fermi energy.  
When this is the case, the superconducting correlations will be 
enhanced.   


\begin{figure}[tb]
$$ \psboxscaled{380}{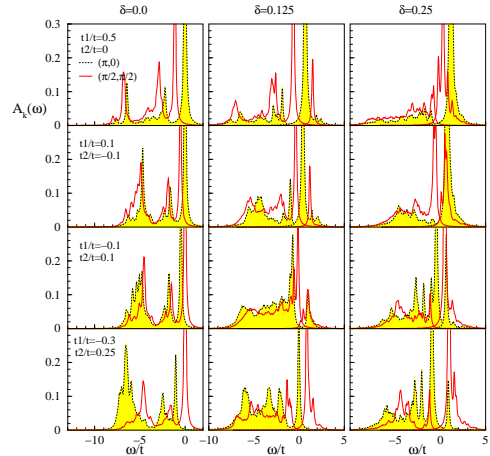} $$ \vspace{-10mm}
\caption{
Single-electron spectral function in the 2D t-J model including 
$t_1$ and $t_2$.  $J/t=0.4$.  
Calculations are performed for all the six 
nonequivalent ${\bf k}$-points, and the results for typical 
values of $t_1$ and $t_2$ are shown here for ${\bf k}$=$(\pi,0)$ 
and $(\pi/2,\pi/2)$.  Particle and hole parts correspond 
to $\omega >0$ and $\omega <0$, respectively.  Fermi energy 
at $\delta=0$ is set that the highest hole peak is at $\omega=0$.  
}
\label{fig2}
\end{figure}

Since the above discussion on the single-electron level 
is within the free electron picture, we have 
to check it directly through the single-electron Green's  
function.  For the same set of parameters as 
in Fig.~1, the spectral function of the Green's function,  
$A_{\bf k}(\omega)$, is calculated and the results are plotted in Fig.~2.   
Fermi energy is here set to be zero, $E_{F}$=0.  
Sharp peaks near $E_F$ are of 
quasiparticle or quasihole, and a broad incoherent background 
is spread at higher $|\omega|$.  
Here, we define the difference in the quasiparticle 
level $u \equiv \varepsilon (\pi,0)- \varepsilon (\pi/2,\pi/2)$, 
from the peak positions in $A_{\bf k}(\omega)$.  
The observed $u$-dependence qualitatively support 
the picture explained before.   
The quasiparticle 
level at ($\pi$,0) shifts downwards 
with decreasing $u_0$, despite the reduction of the 
size of the shift.  
The general trend holds that the ($\pi$,0)-level is 
close to $E_F$ when the pairing 
correlations are enhanced.  
Similar behavior is also observed at other $J/t$'s.  
The reduction of the shift is due to electron-electron 
correlations, since single-electron hopping processes 
lose dispersion and coherence near half filling.  
We note that the t-J model with $t_1$=$t_2$=0 gives 
$u \sim 0$, and this degeneracy also remains in much 
larger systems of the Hubbard model ~\cite{tu0}.

As is mentioned above, the ARPES results \cite{calcium1,calcium2,lsco} 
suggest that the $(\pi,0)$-level is 
substantially lower than the $(\pi/2,\pi/2)$-level in the insulating
or very lightly doped samples. 
From the above argument, if we employ the t-J model,
the deep $(\pi,0)$-level requires that
the distant hopping satisfies $u_0 < 0$ as in
Refs.~\cite{distant,calcium2}.
However, if this is the case, the pairing correlations are not
optimally enhanced at realistic dopings around $\delta \sim 0.15$.
Experimentally, the
$(\pi,0)$-level quickly rises 
up to the $(\pi/2,\pi/2)$-level with further doping.  
This behavior is difficult to explain in the rigid band 
picture, evidence of correlation effects presumably including 
strong pairing interaction.  

\begin{figure}[tb]
$$ \psboxscaled{200}{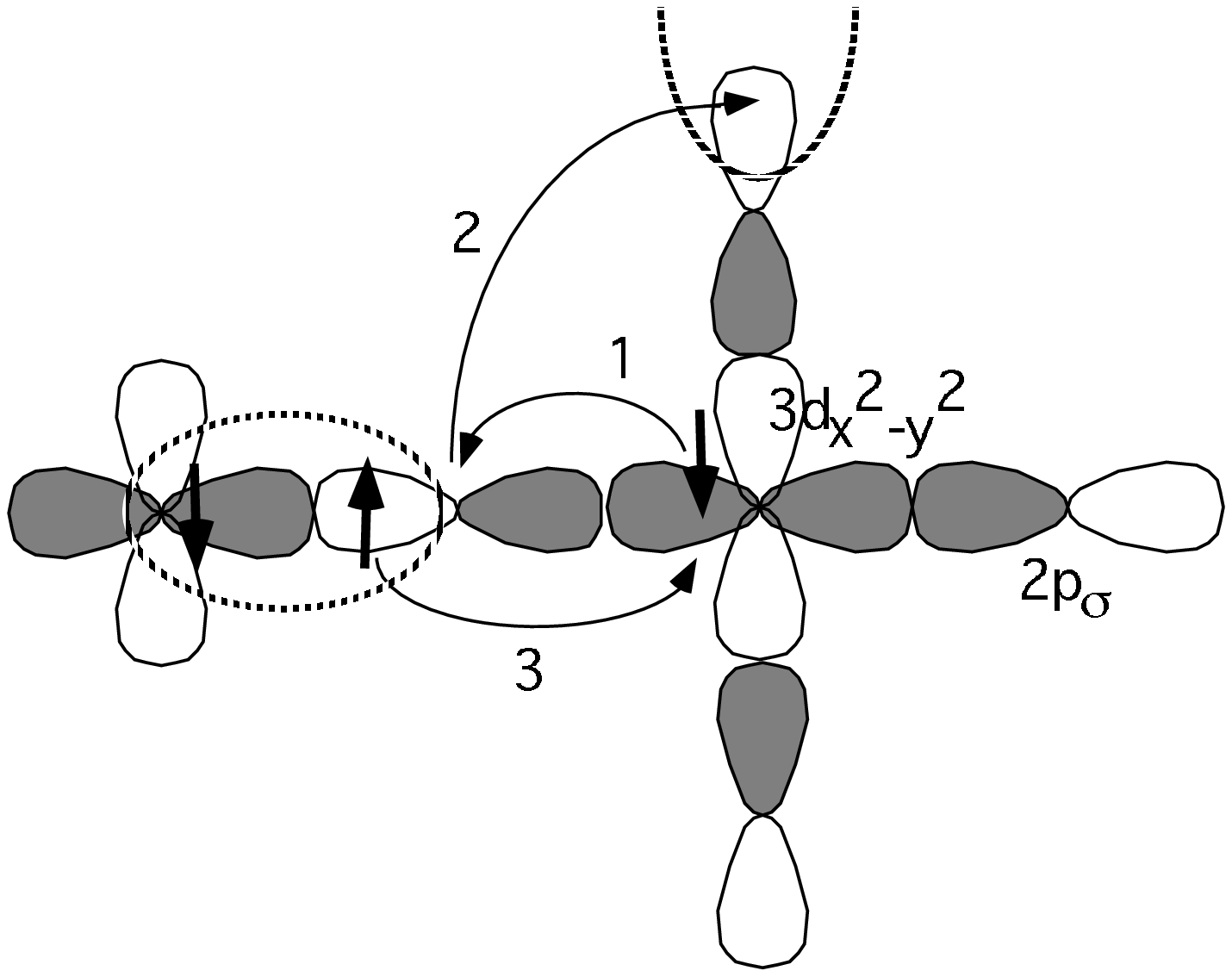} \hspace{0.3cm}
   \psboxscaled{200}{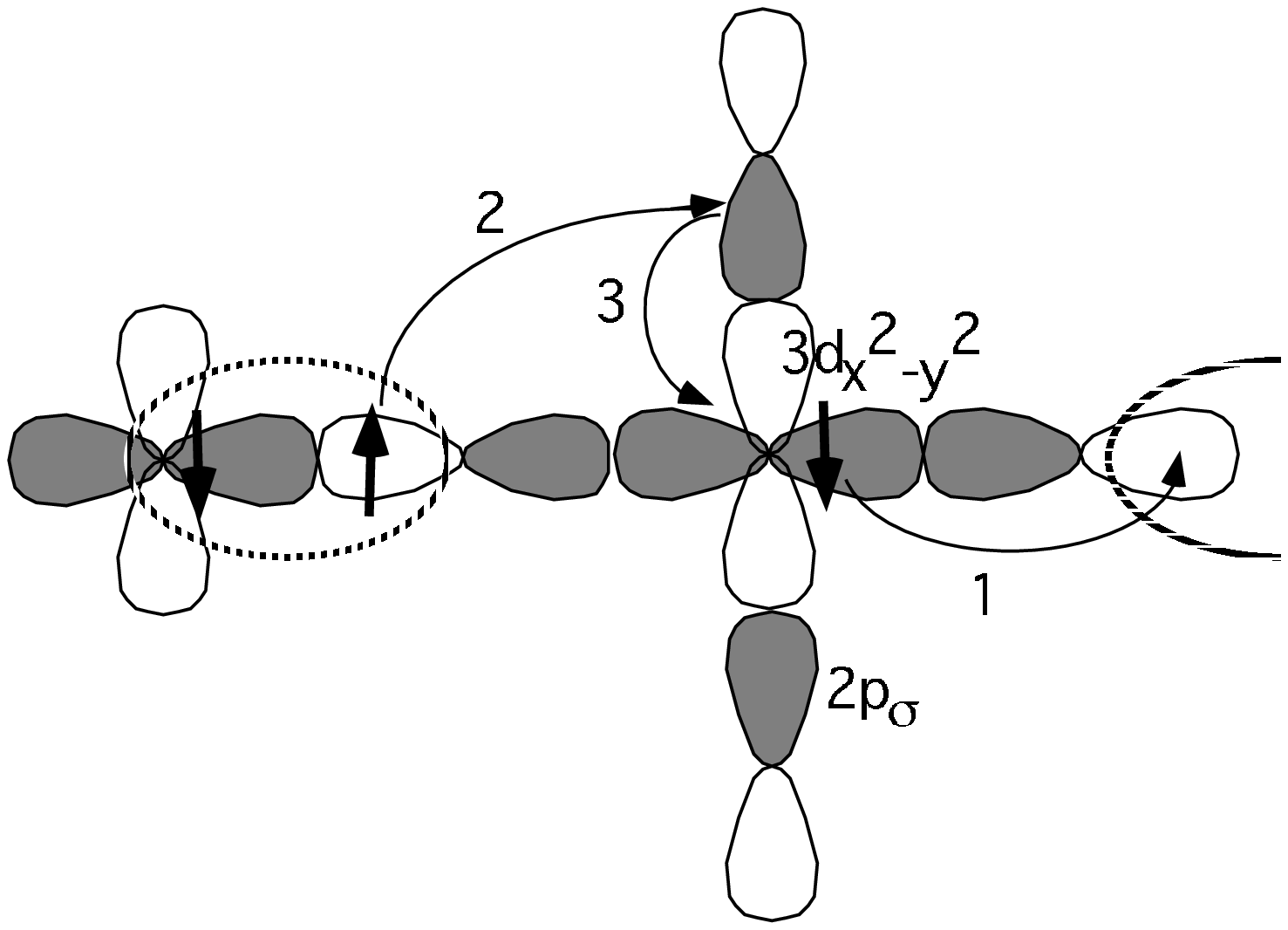} $$ \vspace{-10mm}
\caption{
Two typical third-order terms in the strong 
coupling expansion of the d-p model leading to the W$_d$ term.  
Numbers 1-3 denote the order of the perturbation processes, 
and the process 2 uses the direct O-O hopping.  
Arrows show hole spins, and the Zhang-Rice singlets are 
depicted by dotted ellipses.  
({\em a}): Three-site term with 90$^\circ$ configuration, 
({\em b}): 180$^\circ$ configuration.  
Note that the number of equivalent intermediate configurations 
differs between ({\em a}) and ({\em b}).  
}
\label{fig3}
\end{figure}

To incorporate such pairing effects, 
we now examine microscopic processes which account for generic 
feature of d-wave pairing by employing 
the square of local kinetic energy \cite{tuwd}. 
A part of similar interaction has been studied 
by several authors \cite{tuw,threesite1,threesite2,tjw}, 
but many features remain unexplored.  
The Hamiltonian, say the t-J-W$_d$ model, reads
\begin{equation}
H_{tJW_d} = H_{tJ} -W_d \sum_{j} 
\bigl[ \sum_{a,\sigma}  f_d(a) c_{j\sigma}^\dagger 
 c_{j+a\sigma} + \mbox{H.c.}
\bigr]^2 , 
  \label{tJW}
\end{equation}
where $a$ labels four nearest neighbors of each site, 
and the form factor $f_d (a)$ is $d_{x^2-y^2}$-wave like 
as defined before.  
The $W_d$-term consists of two-site terms and 
three-site terms \cite{tuwd,tuw}.  
The former renormalizes the superexchange coupling constant
as $J_{\rm eff}$=$J+8W_d$, and its effects could be 
already contained in those of the $J$-term.  
On the contrary, the three-site terms have new effects.  
Since it may be rewritten into the form of electron-pair 
hoppings \cite{tuw}, 
it is natural to expect further enhancement of superconductivity 
due to these terms.  
This W$_d$-term may be derived as the low-energy effective 
Hamiltonian starting from the d-p model,\cite{tjw1}
as Zhang-Rice construction of the $J$-term \cite{tjmodel2}, 
but has customarily been omitted 
without examining its relevance.  
One may also obtain the three-site term starting from 
the single-band Hubbard 
model in the limit of strong on-site repulsion \cite{threesite1}, 
but then with a constant form factor, 
$f(\delta) \rightarrow 1$, and its effects have been 
studied by several authors \cite{threesite1,threesite2,tuw,tjw}.  
Generally, starting from the d-p model, the W-term 
with the $d_{x^2-y^2}$-wave form factor, $W_d$, and 
that with the constant form factor, $W_s$, are both obtained. 
Figure 3 sketches two of typical fundamental processes contributing 
to the three-site terms.  The difference of the two contributes 
to the $W_d$-term, while the sum leads to the $W_s$-term.  
Although usually neglected, 
these processes are in fact of lower order than those for 
superexchange, meaning a considerable size of the coupling 
constants, $W_d$ and $W_s$.  

In the following we will demonstrate based on 
numerical calculations an important role of 
the $W_d$-term for the enhancement of superconducting 
correlations. We also show that such pairing interaction
simultaneously induces renormalization of  single-particle dispersion. 
In contrast to the $W_d$-term, the $W_s$-term does not affect 
the single-electron level at ${\bf k}$'s along the Fermi surface 
at half filling \cite{tjw1}.  
Therefore, 
the deep ($\pi$,0)-level observed in experiments 
again require $u_0 <0$ if $W_d=0$ and $W_s \ne 0$, 
and this implies that 
the enhancement of 
superconductivity remains not prominent.  
On the other hand, the $W_d$-term drives pairing instability,  
and simultaneously shifts the single-particle ($\pi$,0)-level downwards 
relative to ($\pi/2$,$\pi/2$) as we show below. 
It turns out that these two phenomena are compatible 
only when the $W_d$-term is playing the driving mechanism 
of pairing.  
Based on these observations, we examine how to optimize 
superconductivity with the $W_d$-term.  

\begin{figure}[tb]
$$ \psboxscaled{450}{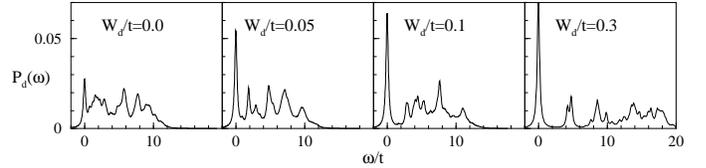} $$ \vspace{-10mm}
\caption{
Dynamic correlations of electron-pair annihilation 
in the 2D t-J-W$_d$ model at $\delta=0$.  
$t_1=t_2=0$ and $J/t=0.3$.  
}
\label{fig4}
\end{figure}

Figure 4 shows the dynamic superconducting correlations 
of the t-J-W$_d$ model (\ref{tJW}), and here $t_1=t_2=0$.  
The growth of the coherent 
peak $P_d (\omega)$ upon switching $W_d$ is noticeable, 
and at the same time the weight of the incoherent background 
at higher energies is substantially suppressed.  Thus 
the effects of the $W_d$-term on the enhancement of superconducting 
correlations are clear.  

The $W_d$-term modifies single-electron dynamics as well.  
Electrons hop using this term to second and third neighbor 
sites with and without spin flip accompanied, 
where the single-electron part of the $W_d$-term 
renormalizes these hoppings as $t_1 + 2W_d$ and $t_2 - W_d$, 
respectively.  
Quasiparticle energies $\varepsilon ({\bf k})$ are 
affected accordingly.  
Of course, electrons are strongly correlated 
near half filling, and we need numerical 
calculations to determine the real quasiparticle energy.   
We have calculated the Green's function  
for the t-J-W$_d$ model now including $t_1$ and $t_2$, 
and determined the hole occupation and 
the quasiparticle/quasihole energy at each ${\bf k}$.  
Some results are shown in Fig.~5.  
As was shown before, the $\delta$-dependence of the single-electron 
level is basically consistent with the rigid band picture, when $W_d =0$.  
On the other hand, for finite $W_d$, 
the single-electron level is strongly renormalized 
to gain a large energy from the $W_d$-term 
rather than single-electron hopping term.  
Its important consequence is the pinning of the ($\pi$,0)-level near 
$E_F$ during hole doping, and the spectral function 
is found to have a 
sharp low-energy peak in both the particle and hole parts.  
This double peak structure agrees with what is expected 
in superconducting states, and it is more prominent than 
that in Fig.~2.  

\begin{figure}[tb]
$$ \psboxscaled{380}{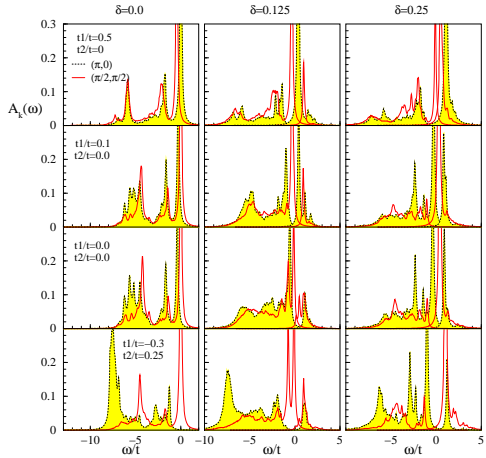} $$ \vspace{-10mm}
\caption{
Single-electron spectral function in the 2D t-W$_d$ model including 
$t_1$ and $t_2$.  $J/t=0$ and $W_d/t=0.05$.  
Results for ${\bf k}$=
$(\pi,0)$ and $(\pi/2,\pi/2)$ are shown here for typical 
values of $t_1$ and $t_2$.  
}
\label{fig5}
\end{figure}

The single-electron part of the $W_d$-term, to some extent, 
suppresses superconductivity near half filling, since 
it pushes down the ($\pi$,0)-level away from $E_F$.  
On more general grounds, 
the d-wave pairing interaction necessarily induces a
downward shift of ($\pi$,0)-level.  
The experimentally observed $(\pi,0)$ level lower than $(\pi/2,\pi/2)$ 
in the highly underdoped samples may at least partially result from
such mechanism.  
We have calculated pairing correlations for the 
t-J-W$_d$ model including $t_1$ and $t_2$ hoppings 
and examined the relation with the ($\pi$,0)-level.  
Figure 6 shows its coherent part $P_d (\omega=0)$ at 
$\delta =0$ as a function of the level difference 
between ($\pi$,0) and ($\pi$/2,$\pi$/2).  
As a realistic choice, we here set 
$J=0$ and $W_d=0.05t$, equivalent to $J_{\rm eff}=0.4t$, 
corresponding to Fig.~1.  
The level difference $u<0$ at half filling and the enhancement 
of the pairing correlations with further doping now becomes 
more compatible.  In addition, as in the t-J model, 
superconductivity is more enhanced when the ($\pi$,0)-level 
is tuned close to the Fermi energy, 
and this indeed offers potential optimization beyond 
present experimental realizations.  

We briefly note an important role of the dynamic exponent $z$ 
of the Mott transition.  It sets the coherence temperature 
as $T^* \sim \delta^{z/2}$,~\cite{scaling,review} 
an essential factor determining $T_c$.  
Scaling of superfluid density is also subject to $z$, 
since its upper bound is given by Drude weight \cite{superfluid} 
and this scales as $D \sim \delta ^{z/2}$ in 2D 
as $\delta \rightarrow 0$~\cite{scaling}.  
Therefore, for larger $z$, the superfluid density vanishes 
faster with approaching half filling. 
When the W$_s$-term is switched on, we have found the 
change in $z$ from the unusual value $z$=4 to 2~\cite{tjw1}, 
indicating an incoherent-to-coherent transition.  
We speculate that the enhancement of superconductivity 
we observed near half filling is also associated with 
this transition of $z$ caused by the presence 
of the $W_d$-term together with the flat band around 
$(\pi,0)$ pinned near $E_F$.  
We will discuss this point in more detail in a future publication.


In conclusion, 
the flat dispersion near $(\pi,0)$ is determined 
self-consistently from strong correlations 
and pairing effects, and 
the d-wave superconducting correlation 
is enhanced when the resultant band is pinned near the Fermi level.  
Because of its large dependence, it opens a possibility 
of further optimization of 
pairing correlation, and hence a chance to achieve 
a higher transition temperature by careful tuning. 
In the light of this level tuning, the high-Tc cuprates
may not be in the optimized condition, because of the
apparently lower level at $(\pi,0)$ than $(\pi/2,\pi/2)$.  
This is presumably due 
to the same origin with the pairing force, and there 
is a good chance for further tuning.
Our proposal is that the bare single-particle transfer 
should be chosen to keep the renormalized flat dispersion 
near the Fermi level by compensating the downward shift of 
the ($\pi$,0)-level induced by the d-wave pairing interaction itself.  
This offers a typical way to achieve the enhancement of 
superconductivity by a careful tuning of the lattice
parameters.  
For the quest of new materials, the bare single-particle level
suggested from the band structure calculation would
be helpful if combined with our procedure.

\begin{figure}[tb]
$$ \psboxscaled{380}{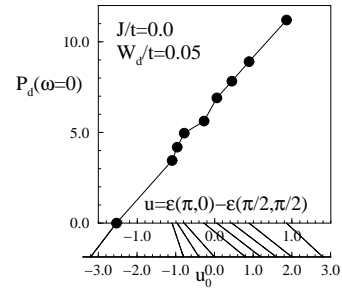} $$ \vspace{-10mm}
\caption{
Enhancement of superconductivity by level tuning 
in the 2D t-W$_d$ model. 
The coherent part of 
electron-pair annihilation at $\delta=0$ is plotted as a 
function of the level difference $u$ between 
$(\pi,0)$ and $(\pi/2,\pi/2)$.  Level difference for the 
free electron case $u_0$ is also shown for comparison.  
}
\label{fig6}
\end{figure}

The authors thank Fakher Assaad for valuable discussions.  
This work is supported by a Grant-in-Aid for 
``Research for the Future'' Program from Japan Society for 
the Promotion of Science under the project 
JSPS-RFTF97P01103.

\end{document}